\documentclass[runningheads]{llncs}
\usepackage{graphicx}
\usepackage[table,x11names]{xcolor}
\usepackage{multirow}
\usepackage{caption}
\usepackage{subcaption}

\newcommand \ignore[1]{}
\newcommand*{\affaddr}[1]{#1} 
\newcommand*{\affmark}[1][*]{\textsuperscript{#1}}

\begin{document}
\title{Incremental Transfer Learning: An Approach to Detect COVID-19 Coughs from Healthy People's Cough Detection Models and Fewer Patient Coughs}
\title{Incremental Transfer Learning to Detect COVID-19 Coughs from Healthy People's Cough Detection Models and Fewer Patient Coughs}
\title{Detect COVID-19 Coughs from Healthy People's Cough Models and Folds of Patient Coughs Using Transfer Learning}
\title{Detect COVID-19 Coughs from Healthy People's Cough Models by Incrementally Adding Patient Cough-Folds with Transfer Learning}
\title{Transfer Learning to Detect COVID-19 Coughs from Healthy People's Cough Models and Incremental Addition of Patient Coughs}
\title{Transfer Learning to Detect COVID-19 Coughs with Incremental Addition of Patient Coughs to Healthy People's Cough Detection Models}
\titlerunning{Transfer Learning}
%

\author{
Sudip Vhaduri\affmark[1], Seungyeon Paik\affmark[1], and Jessica E Huber\affmark[2,3]\\
\affaddr{\affmark[1]Computer and Information Technology Department\\ 
\affmark[2]Speech, Language, and Hearing Sciences Department\\
Purdue University, West Lafayette, IN 47907, USA}\\
\affmark[3]Communicative Disorders and Sciences Department\\
University at Buffalo, Buffalo, NY, 14214, USA\\
\email{\affmark[1]\{svhaduri,paiks\}@purdue.edu\\
\affmark[2]jhuber@purdue.edu,,\affmark[3]jehuber@buffalo.edu}
}
\authorrunning{S. Vhaduri \emph{et al.}}
\institute{}

\maketitle              
\begin{abstract}
Millions of people have died worldwide from COVID-19. In addition to its high death toll, COVID-19 has led to unbearable suffering for individuals and a huge global burden to the healthcare sector. Therefore, researchers have been trying to develop tools to detect symptoms of this human-transmissible disease remotely to control its rapid spread. 
Coughing is one of the common symptoms that researchers have been trying to detect objectively from smartphone microphone-sensing.
While most of the approaches to detect and track cough symptoms rely on machine learning models developed from a large amount of patient data, this is not possible at the early stage of an outbreak. 
In this work, we present an incremental transfer learning approach that leverages the relationship between healthy peoples' coughs and COVID-19 patients' coughs to detect COVID-19 coughs with reasonable accuracy using a pre-trained healthy cough detection model and a relatively small set of patient coughs, reducing the need for large patient dataset to train the model.
This type of model can be a game changer in detecting the onset of a novel respiratory virus.  
\keywords{transfer learning \and COVID-19 \and Cough detection.}
\end{abstract}
%
%
%


\section{Introduction}\label{introduction}

\subsection{Motivation}

Coronavirus and other human-transmissible respiratory viruses have become prevalent and have led to human suffering and a large number of deaths in recent times. 
According to the World Health Organization (WHO), the novel coronavirus SARS-CoV-2 (COVID-19) has so far caused a total of over 771 million infections and over 6.9 million deaths globally~\cite{who}. Even after the development of a vaccine, over 300 thousand infections and 1.5 thousand deaths occur a day~\cite{who}. 
Additionally, COVID-19 created a heavy economic burden on the health sectors, e.g., the United States incurred a total of \$163.4 billion in direct medical expenses during the pandemic~\cite{richards2022economic}.
Early onset detection can help prevent the rapid spread and its adverse consequences. 
But, traditional diagnosis approaches are slow and require resources, such as viral tests (based on samples from the nose and mouth) or antibody tests~\cite{Testcovid19_cdc}, chest X-ray or spirometry tests~\cite{COPD_treatment}, blood tests, pulse oximetry, and sputum tests~\cite{Pneumonia,Asthmadiagnose}. These resources are not readily available in peoples' homes or at healthcare access points, such as primary care or urgent care. Therefore, there is a need for an approach that can be easily deployed to quickly detect the onset and control disease spread. 


\subsection{Related Work}\label{relatedWork}

Researchers have been trying to develop tools/systems to objectively detect and remotely report typical symptoms of respiratory diseases, such as coughing. Many of these techniques require the use of wearable technology.
For example, researchers have detected coughing with 0.82 accuracy using smartwatch accelerometers and audio recordings~\cite{liaqat2021coughwatch}, and 0.94 -- 0.95 sensitivity using ECG, thermistor, chest belt, accelerometer, contact microphone, audio microphone~\cite{drugman2013objective} and chest sensor ~\cite{amoh2015deepcough}.
%
Some researchers proposed a respiratory monitoring system using a wearable patch sensor~\cite{elfaramawy2018wireless} and a wearable radio-frequency (RF) cough monitoring system~\cite{hui2021wearable}.
On the other hand, a group of researchers proposed a COVID-19 symptom tracker utilizing a headset-like sensor~\cite{stojanovic2020headset}.

However, we have found that people's adherence to wearables drops significantly over time, compared to smartphone adherence~\cite{wearablecons,kamei2022use}. 
Therefore, some researchers have been trying to detect objective symptoms, such as coughing, using smartphone data~\cite{vhaduri2023environment,dibbo2021effect,chang2022covnet,vhaduri2019towards,cai2023discovering}. One team of researchers proposed machine learning-based COVID-19 cough, breath, and speech detection using smartphone recording files~\cite{pahar2021machine,pahar2021covid}. They achieved up to 0.93 area under the curve using the k-nearest neighbor classifiers. Other researchers developed frameworks to diagnose COVID-19 using a smartphone app and built-in sensors in the smartphone~\cite{imran2020ai4covid,laguarta2020covid}.

However, a major limitation of all these existing cough detection models is the underlying assumption of the availability of a good amount of relevant data, which is not always possible~\cite{chang2022covnet}. For example, during the early stage of a new outbreak, there is not much data to develop a reasonably good traditional machine learning model due to the need for a large volume of data. But coughs from healthy people and patients have similarities, which can be utilized to detect COVID-19 coughs using a healthy cough detection model and a relatively small set of coughs from COVID-19 patients, reducing the need for large amounts of COVID-19 patient coughs. This kind of model would be invaluable in the detection of a new novel respiratory disease.

\subsection{Contribution}

\textcolor{black}{The main contribution of this work is to present a novel approach leveraging similarities between healthy people's coughs and COVID-19 patients' coughs to incrementally transfer healthy cough detection models to COVID-19 cough detection models with smaller batches or folds of COVID-19 coughs.
Compared to large data-driven traditional modeling approaches, incremental transfer learning approaches can help detect the onset of a novel respiratory virus early utilizing a relatively small set of cough samples obtained from the first few people infected and a pre-trained healthy people's cough detection model to control the spread of the novel respiratory disease to minimize adverse consequences.}

\begin{figure}
\centering
\includegraphics[width=0.95\linewidth]{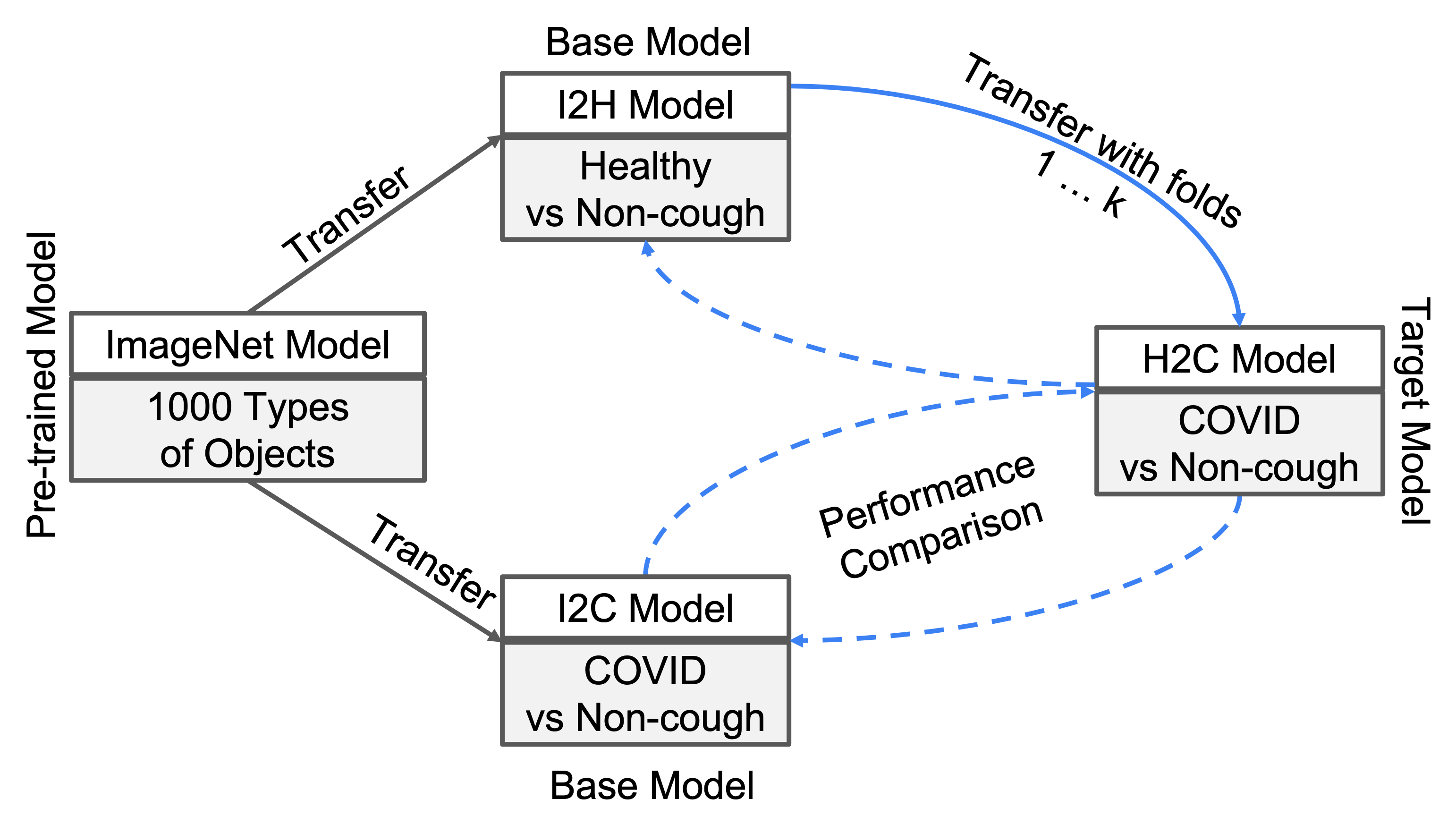}
\caption{\textcolor{black}{Proposed modeling approach}}
\label{overview}
\end{figure}


\section{Methods}\label{method}

In this section, we first present our modeling approach. Next, we discuss cough and non-cough audio recordings obtained from different sources, data processing steps, and neural network architectures and parameters used in this work.  

\subsection{Modeling Approaches}\label{approaches}

In Figure~\ref{overview}, we present our modeling scheme. 
We start with a pre-trained model~\cite{simonyan2014very} developed to detect 1000 objects, i.e., classes from the ImageNet dataset~\cite{imagenet}. 
\textcolor{black}{Next, we use transfer learning to develop two base models and one target model to detect coughs obtained from healthy people and COVID-19 patients using transfer learning. 
\begin{enumerate}
    \item ImageNet to Healthy ({\bf I2H}) model: We transfer our ImageNet 1000 object detection model to a binary model detecting healthy cough versus non-cough. Throughout this manuscript, we name this model as ``ImageNet to healthy'', i.e., {\bf I2H} model. This is one of the two base models. This model will be later used to develop the target COVID-19 cough detection model incrementally. 
    \item ImageNet to COVID-19 ({\bf I2C}) model: We transfer the ImageNet 1000 object detection model to a binary model that detects COVID-19 cough versus non-cough, which we refer to as ``ImageNet to COVID-19'', i.e., {\bf I2C} model in this manuscript. This is the second base model, which will be used as a reference model when comparing the performance of our target COVID-19 cough detection models. 
    \item Healthy to COVID-19 ({\bf H2C}) model: We transfer the healthy cough detection model, i.e., {\bf I2H} model, incrementally by adding smaller batches, i.e., folds of COVID-19 coughs to detect COVID-19 coughs. This target model is named ``Healthy to COVID-19'', i.e., {\bf H2C} model.
\end{enumerate}
}

\textcolor{black}{Our ultimate goal is to utilize the {\bf I2H} model and smaller batches/folds of COVID-19 coughs to develop a target {\bf H2C} model that achieves close performance to the {\bf I2C} base model to investigate the capability and feasibility of incremental transfer learning.}
\textcolor{black}{The incremental fold addition is the core of our transfer learning approach. Model performance is analyzed to show how the target models improve over time and to determine the minimum number of samples required to identify disease coughs accurately. There may not be a base model for a new disease for the performance comparison, such as the {\bf I2C} model. However, our approach is to demonstrate that incremental transfer learning can be a pathway to get a valid model even when we have few samples of the new disease-specific patient data.} 



\begin{figure*}
\centering
\includegraphics[width=.95\linewidth]{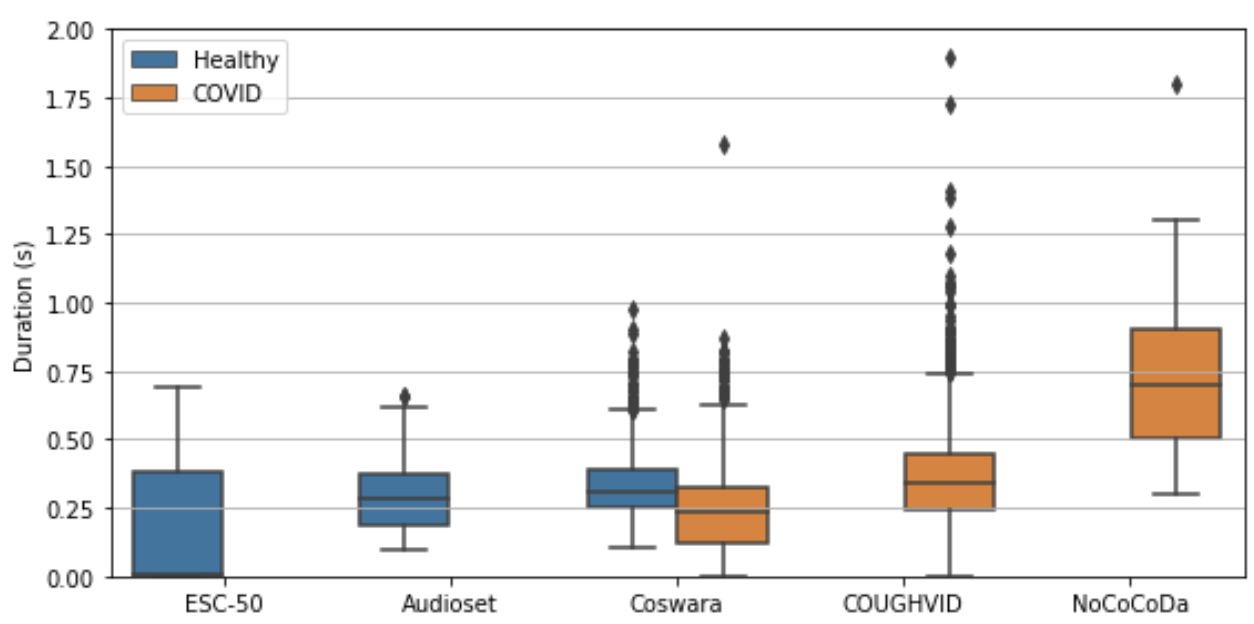}
\caption{Distribution of cough duration}
\label{audio_length}
\end{figure*}

\subsection{Datasets}\label{datasets}

This section briefly describes six major datasets used in our experiments, including three COVID-19, two healthy cough datasets, and an image dataset. 

\begin{enumerate}
    \item ImageNet Dataset~\cite{imagenet}: This is a publicly available image dataset with a thousand object classes and millions of images. The dataset contains various types of objects types, including geese, balloons, and fruits. We use the ImageNet dataset to develop the pre-trained ``ImageNet'' model. 
    \item Coswara Dataset~\cite{sharma2020coswara}: This crowdsourced dataset consists of breathing, coughing, and voice sound recordings from healthy people and COVID-19 patients. The sampling rate is in the range of 47.82 $\pm$ 0.83 kHz. In this work, we use the voluntary coughs obtained from 274 COVID-19-positive patients.  
    \item COUGHVID Dataset~\cite{orlandic2021coughvid}: This crowdsourced dataset contains more than 25,000 cough recordings from COVID-19-positive patients or asymptomatic people of varying ages, genders, and regions. The sampling rate of this dataset is around 44.1 kHz. We use the voluntary coughs obtained from 719 COVID-19-positive patients or asymptomatic people.
    \item NoCoCoDa Dataset~\cite{cohen2020novel}: This dataset contains natural cough recordings of 13 COVID-19-positive patients collected from public media interviews. Audio recordings are collected at a 44.1 kHz sampling rate. 
    \item ESC-50 Dataset~\cite{piczak2015esc}: The environmental sound classification (i.e., ESC-50) dataset consists of audio recordings from five categories of sounds (i.e., animal, natural soundscapes, human sounds, interior sounds, and exterior noises) with 10 types of sounds per category recorded at a rate of 44.1 kHz. There are 40 audio recordings per type (2,000 recordings in total).
    \textcolor{black}{We use the voluntary cough recordings from five healthy subjects as the cough class, i.e., healthy coughs. The remaining 49 sound types are used to create the non-cough class when developing the binary cough versus non-cough models presented in Section~\ref{approaches}.}
    \item AudioSet Dataset~\cite{gemmeke2017audio}: This dataset contains a wide range of 632 sound classes obtained from YouTube videos, where samples are recorded at 16 kHz and 44.1 kHz. In this work, we use voluntary coughs from 88 healthy subjects. 
\end{enumerate}


In this work, we combined all COVID-19 data to create one COVID-19 patient cough dataset (n = 1006 patients). Similarly, we combined all healthy cough datasets to develop another cough dataset (n = 83 healthy people). While combining data from different datasets, we keep the subject information so that they can be utilized later to create mutually exclusive splits or folds among the training, validation, and test sets. All non-cough data is obtained from the ESC-50 dataset.


\subsection{Data Processing}\label{pre-proc}

This section presents our cough ground-truth label collection approach from long audio recordings, followed by additional processing steps, including finding an optimal window size, padding, and feature extraction. Finally, we present our cross-validation approach with incremental training to develop {\bf H2C} models from {\bf I2H} model which requires relatively fewer COVID-19 coughs. 

\subsubsection{Data Cleaning and Ground-Truth Label Collection}\label{labeling}

Most of the audio data used in this work come from crowd-sourced datasets. Therefore, we first perform a rigorous cleaning process to remove unwanted parts, including quiet, speech, and noises using different audio signal processing libraries and tools, including the Audacity toolbox~\cite{audacity}. 

After the initial cleaning, we performed data segmentation to extract the ground-truth cough labels from long audio recordings with multiple coughs utilizing the Audacity toolbox. 
Adapted from our previous work~\cite{vhaduri2019nocturnal,vhaduri2020nocturnal}, we automated the process by developing an energy threshold-based audio segmentation followed by a phase classification approach.

We obtained 144 and 252 healthy coughs from the ESC-50 and AudioSet datasets. Similarly, we obtained 1892, 2690, and 73 COVID-19 coughs from the Coswara, COUGHVID, and NoCoCoDa datasets, respectively. 
In total, we used 396 healthy coughs and 4655 COVID-19 coughs.

\begin{figure*}
\centering
\begin{subfigure}{.95\textwidth}
  \centering
  \includegraphics[width=.8\linewidth]{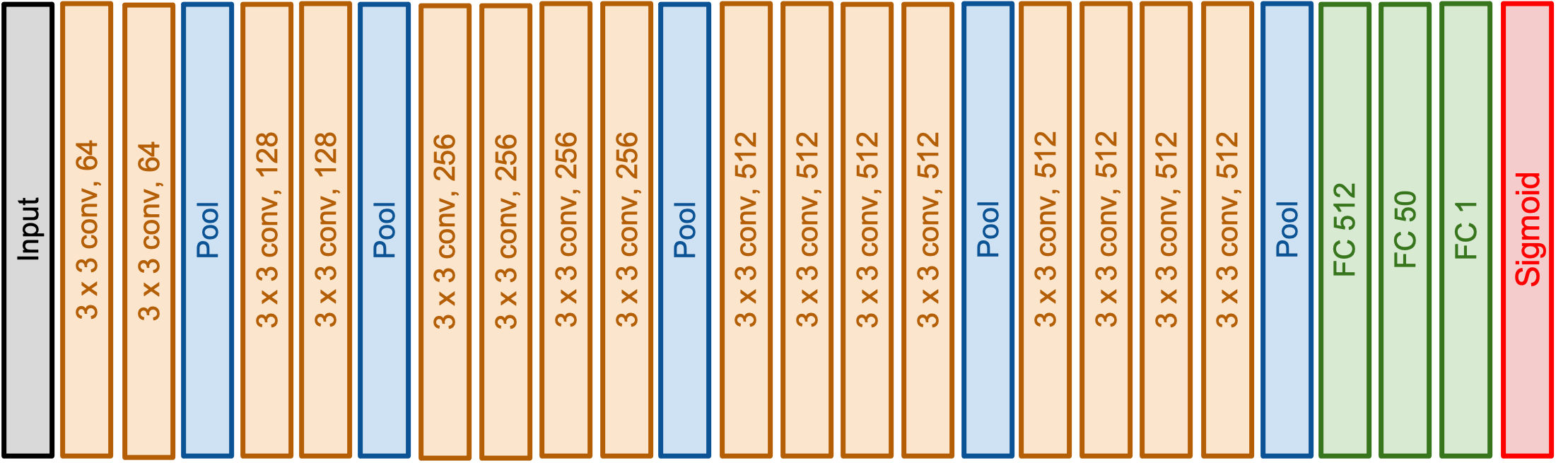}
  \caption{I2H or I2C Models}
  \label{I2H_I2C}
\end{subfigure}
\begin{subfigure}{.95\textwidth}
  \centering
  \includegraphics[width=1.0\linewidth]{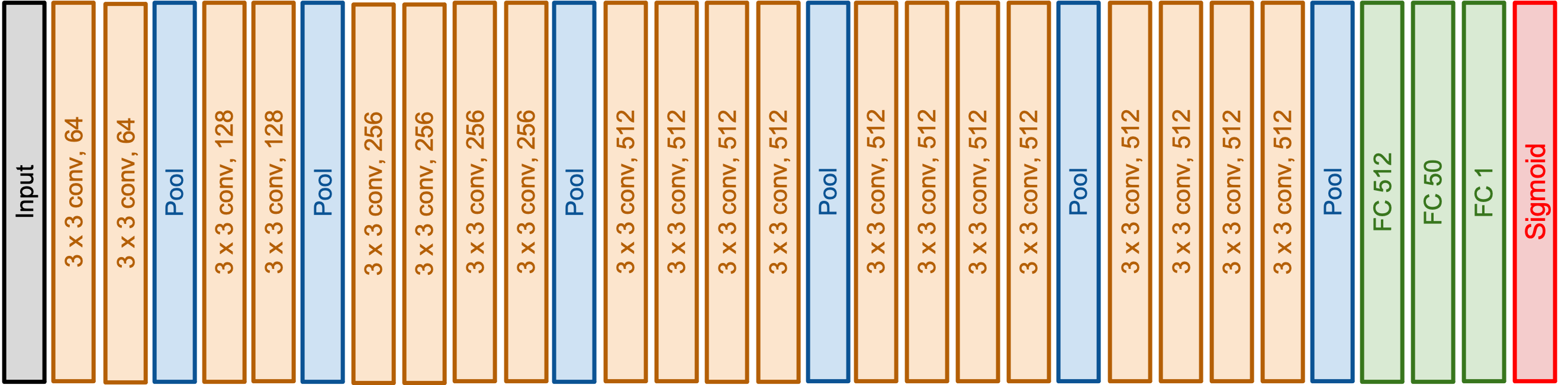}
  \caption{H2C Model}
  \label{H2C}
\end{subfigure}
\caption{\textcolor{black}{Optimal Architecture of different models}}
\label{architecture}
\end{figure*}

\subsubsection{Feature Extraction}\label{feature}

Since we collected audio recordings from different sources, we first changed all sampling frequencies to 44.1 kHz frequency before additional processing. 

Next, we determined a suitable window size before computing features. In Figure~\ref{audio_length}, we use boxplots to represent the distribution of cough duration obtained from various datasets. In the figure, we find that $75^{th}$ percentile of Coswara healthy cough duration, i.e., 0.3917 seconds, is a suitable choice for window size since most of the other healthy and patient coughs, except the NoCoCoDa coughs, have a duration shorter than that. Compared to other datasets, the NoCoCoDa dataset has relatively fewer samples.  

We perform padding (i.e., add 0s) at the end of coughs shorter than the window size (i.e., 0.3917 seconds). In the case of longer cough and non-cough audio recordings, we truncate the parts longer than the window size. 
Finally, we compute the log mel-spectrogram~\cite{meng2019speech} (logmel) \textcolor{black}{and Mel-frequency cepstral coefficient (mfcc)} image features from each cough and non-cough audio recording using the Python Librosa library~\cite{mcfee2015librosa}.



\subsubsection{Training, Validation, and Test Split}\label{splits}

First, we uniformly split the cough data into 10 mutually exclusive folds based on subjects. 
\textcolor{black}{For class balancing, we select the same number of non-cough samples as we have cough samples for the 10 folds. Thereby, we maintain the same number of cough (either healthy or COVID-19) and non-cough instances when training, validating, and testing binary cough (either healthy or COVID-19) versus non-cough detection models presented in Section~\ref{approaches}.}
Next, we follow a ``leave-2-fold-out'' for validation and ``leave-2-fold-out'' for the test approach, where we use the remaining six folds for training while developing and validating/testing a {\bf I2H} or {\bf I2C} model from the pre-trained ImageNet Model. We follow a rotational approach, where we perform this mutually exclusive 6-2-2 train-validation-test split 10 times to develop 10 different models.
While developing {\bf H2C} models, we follow an incremental training approach, where we add the six training folds of COVID-19 data to the {\bf H2C} models one-by-one. 
We perform this incremental training for one of the 10 6-2-2 splits.
\textcolor{black}{During the incremental training, we distribute COVID-19 subjects in 10 folds using a snaking approach, where we first sort the subjects in descending order of cough counts. Then, we put the top 10 subjects into 10 folds. Next, 10 subjects are put into 10 folds. This way, we maximize the number of cough samples in 10 folds. We end up with 100 subjects in each fold with 450 random COVID-19 cough samples.}

\subsection{Neural Network Architectures and Parameters}

\textcolor{black}{In Figure~\ref{I2H_I2C}, we present the optimal architecture of the two models (i.e., {\bf I2H} or {\bf I2C} models) transferred from the pre-trained ImageNet model developed with VGG19 ~\cite{simonyan2014very}. In Figure~\ref{H2C}, we present the optimal architecture of the {\bf H2C} model transferred from the {\bf I2H} model by adding folds of COVID-19 coughs. 
In the figure, the sequence and meaning of different parameters in each layer are kernel size and the number of feature maps, i.e., nodes in each layer.}
The pre-trained ImageNet VGG19 model had fully connected (FC) three layers with 4096, 4096, and 1000 nodes, followed by a softmax layer to classify 1000 objects. We changed the last four layers with FC 512, 50, and 1 node, followed by a sigmoid layer for binary classification of healthy cough versus non-cough ({\bf I2H} model) or COVID-19 cough versus non-cough ({\bf I2C} model). 
\textcolor{black}{Compared to the {\bf I2H} or {\bf I2C} model, we add four additional convolutional layers and a pooling layer in the case of {\bf H2C} models.}

We used TensorFlow and Keras libraries to develop our models.
We used the ReLU activation function in the hidden layers and the sigmoid decision function in the final layer for every model. For the loss function, we used binary cross-entropy. We tried Adam and RMSprop optimizer and found Adam achieves 53\% higher accuracy than the RMSprop. We also tried batch sizes 16 and 32 and found that batch size 16 is more accurate and has a shorter execution time. The input size used in this work is (320,320). We considered a range of learning rates, including 0.00001, 0.0001, and 0.0005, and epochs ranging from 30 to 200. 
We found a learning rate of 0.0001 works better for the {\bf I2H} and {\bf I2C} models. Similarly, we found a learning rate of 0.00001 is a good compromise for the {\bf H2C} models.
We developed models on Purdue University's Gilbreth GPU server with 8 GPU nodes and 16 cores per node~\cite{gilbreth}. Each node has 192 GB memory, 100 Gbps Infiniband interconnects, and 2 P100 GPUs.


\begin{figure}
\centering
\includegraphics[width=.7\linewidth]{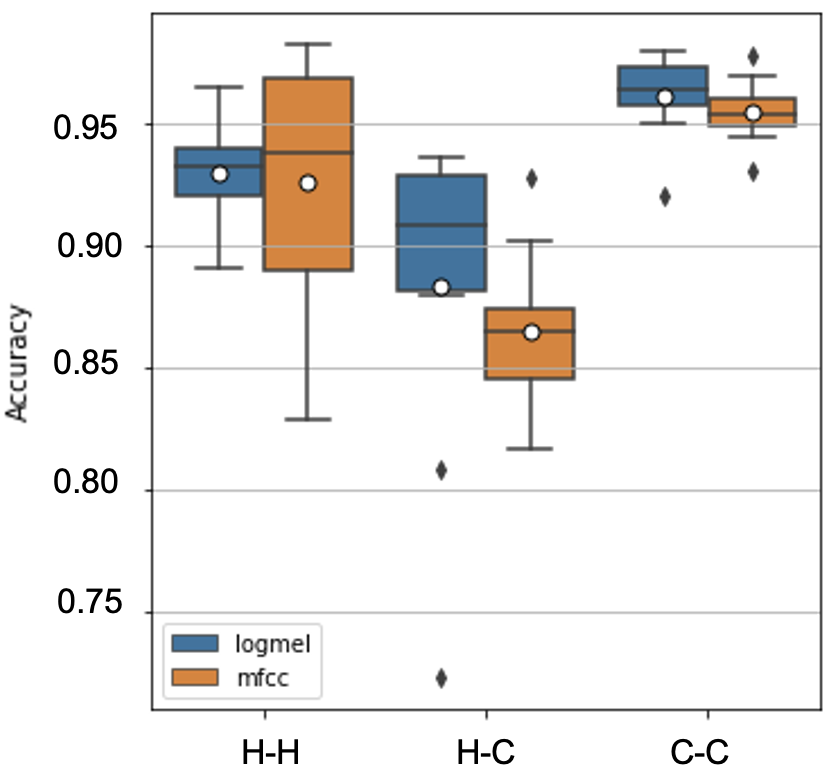}
\caption{\textcolor{black}{{\bf I2H} and {\bf I2C} models tested on different coughs (H-H refers to the case when {\bf I2H} models are tested on healthy coughs; similarly, H-C and C-C refer to the cases when {\bf I2H} and {\bf I2C} models are tested on COVID-19 coughs)}}
\label{I2H-I2C-Performance}
\end{figure}

\begin{figure*}
\centering
\includegraphics[width=.95\linewidth]{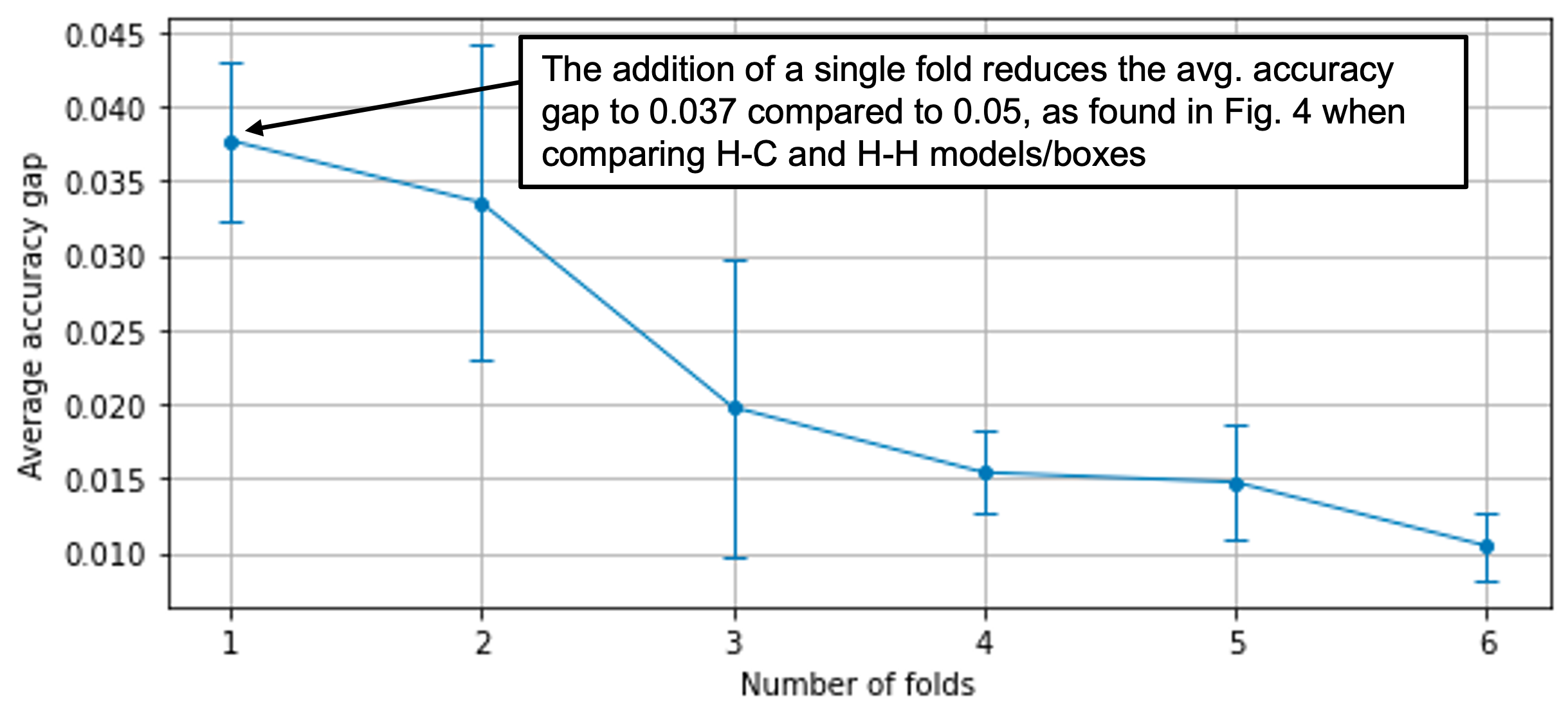}
\caption{\textcolor{black}{Performance of {\bf H2C} models developed using transfer learning incrementally by adding folds of COVID-19 coughs to {\bf I2H} models}}
\label{H2C-Performance}
\end{figure*}

\section{Results}\label{result}

\textcolor{black}{Since we use the mutually exclusive folds with the same number of cough and non-cough instances in each fold (Section~\ref{splits}) when developing binary models, our classes are always balanced, and we use classification accuracy when comparing the performance of different models.}

In Figure~\ref{I2H-I2C-Performance}, we present the performance of the two base models, i.e., {\bf I2H} and {\bf I2C} models when tested on different cough datasets. 
First, we analyze the performance of the {\bf I2H} base models, i.e., healthy cough detection models transferred from the pre-trained ImageNet models using healthy people's coughs. The {\bf I2H} base models are tested on the healthy test coughs, and the findings are presented using the Healthy-Healthy or H-H box in the figure. 
\textcolor{black}{We observe average accuracy values of 0.93 $\pm$ 0.02 (logmel) and 0.92 $\pm$ 0.05 (mfcc) with median values of 0.93 (logmel) and 0.93 (mfcc).}

Next, we analyze the performance of the second base model, i.e., {\bf I2C} model (COVID-19 cough detection models transferred from the pre-trained ImageNet models using COVID-19 patient coughs). The {\bf I2C} model is tested on the COVID-19 test coughs, and the findings are presented using the COVID-COVID or C-C box in the figure. 
\textcolor{black}{We observe average accuracy values of 0.96 $\pm$ 0.02 (logmel) and 0.95 $\pm$ 0.01 (mfcc) with median values of 0.96 (logmel) and 0.95 (mfcc).}

Next, we analyze the performance of the {\bf I2H} models (i.e., base models to detect healthy people's coughs) when tested on COVID-19 test coughs and present the findings using the Healthy-COVID or H-C box in the figure. 
\textcolor{black}{We observe average accuracy values of 0.88 $\pm$ 0.07 (logmel) and 0.86 $\pm$ 0.03 (mfcc) with median values of 0.91 (logmel) and 0.86 (mfcc).} 
In the case of logmel feature, the drop in average accuracy is 0.05 and 0.08 when compared with the findings in the Healthy-Healthy (H-H) and COVID-COVID (C-C) boxes, respectively. 
\textcolor{black}{In the case of mfcc feature, the drop in average accuracy is 0.06 and 0.09 when compared with the findings in the Healthy-Healthy (H-H) and COVID-COVID (C-C) boxes, respectively.}
To improve the accuracy values of the H-C box (i.e., {\bf I2H} models tested on COVID-19 test coughs), we incrementally developed {\bf H2C} models transferring the {\bf I2H} models by adding small amounts of COVID-19 coughs in folds. 
\textcolor{black}{Since logmel features outperform the mfcc features across all measures, we consider the logmel features in the next analysis.}

\textcolor{black}{In Figure~\ref{H2C-Performance}, we present the accuracy gap of {\bf H2C} models with respect to the average accuracy of the base {\bf I2C} models by varying amounts of COVID-19 coughs (i.e., number of folds) included in the training set. In the figure, we observe that with the addition of only one fold of COVID-19 coughs, the average accuracy gap drops to 0.037.}

\textcolor{black}{The average accuracy gap drops to 0.02 by adding two more folds of COVID-19 coughs to the base {\bf I2H} models. Thereby, with the addition of three folds of COVID-19 coughs with the healthy cough detection models, i.e., {\bf I2H} models, we can achieve a performance close to that of the base COVID-19 detection model performance.}

\textcolor{black}{As we continue adding more folds of COVID-19 coughs to the base {\bf I2H} models, we witness a drop in accuracy gap, and after adding all six folds of COVID-19 patient coughs, the accuracy gap drops to 0.01. Additionally, adding more folds makes the error bar tighter, reflecting more consistent accuracy values. Thereby, using this incremental transfer learning approach, we can develop COVID-19 detection models, such as {\bf H2C} models, from the base healthy people's cough detection models, i.e., {\bf I2H} models and smaller amounts of COVID-19 coughs to achieve very close performance to that of the base COVID-19 detection models, i.e., {\bf I2C} models.}

\section{Conclusion and Discussion}\label{Conclusion}

This work attempts to utilize the power of transfer learning and similarities between two types of coughs, i.e., healthy and COVID-19 coughs, to incrementally develop new models requiring a relatively small set of patient coughs to achieve similar performance to that of the COVID-19 cough detection models trained from a bigger COVID-19 cough dataset. Our findings show the promise of utilizing healthy cough detection models to detect COVID-19 coughs after training with relatively fewer patient coughs. 

This model can be useful to detect an early-onset novel respiratory virus with a smaller amount of relevant data. However, before generalizing the findings to similar or other problems, extended studies with a diverse population, diseases, and stages will be needed. 
While image feature-based transfer learning has been adopted in this feasibility work, in the future, other types of data, e.g., acoustic signals, can be utilized to adapt transfer learning models and can be compared with this feasibility work.
This work and findings will also impact other domains of predictive modeling, including place of importance discovery~\cite{vhaduri2016cooperative,vhaduri2017discovering,vhaduri2018hierarchical,vhaduri2018opportunisticICHI,vhaduri2018opportunisticTBD}, health condition monitoring~\cite{sharmin2015visualization,vhaduri2018impact,vhaduri2016assessing,chen2020estimating,vhaduri2022understanding,simhadri2022understanding,vhaduri2020adherence} and well-being tracking~\cite{vhaduri2014estimating,vhaduri2016human,vhaduri2015design,vhaduri2017design,dibbo2021visualizing,vhaduri2021predicting,kim2020understanding,vhaduri2021deriving,vhaduri2023predicting}, securing a user's cyber-physical space~\cite{vhaduri2019multi,vhaduri2017wearable,vhaduri2018biometric,muratyan2021opportunistic,cheung2020context,vhaduri2017towards,dibbo2021onphone,cheung2020continuous,vhaduri2021HIAuth,vhaduri2022predicting,vhaduri2019summary,vhaduri2023bag,lien2023challenges,al2009load,vhaduri2023implicit,dibbo2023sok,dibbo2023lcanets++}, as it presents the feasibility of developing predictive models with relatively small datasets to alternate the traditional approaches requiring large-scale datasets.

\section*{Acknowledgement}
The authors would like to thank the Clifford B. Kinley Trust for funding this research. 

\bibliographystyle{splncs04}
\bibliography{reference1}

\begin{thebibliography}{10}
\providecommand{\url}[1]{\texttt{#1}}
\providecommand{\urlprefix}{URL }
\providecommand{\doi}[1]{https://doi.org/#1}

\bibitem{Asthmadiagnose}
{Asthma: Steps in testing and diagnosis - Mayo Clinic}. Available: \url{https://mayocl.in/3vPs3J7} (Accessed: January 2023), [Online]

\bibitem{audacity}
{Audacity: Free, open source, cross-platform audio software}. Available: \url{https://www.audacityteam.org/} (Accessed: January 2023), [Online]

\bibitem{Testcovid19_cdc}
{CDC: COVID-19 Testing}. Available: \url{https://bit.ly/3nRjYOM} (Accessed: January 2023), [Online]

\bibitem{COPD_treatment}
{COPD Symptoms and Diagnosis | American Lung Association}. Available: \url{https://bit.ly/3hefi2f} (Accessed: January 2023), [Online]

\bibitem{gilbreth}
{Gilbreth}. Available:\url{https://www.rcac.purdue.edu/compute/gilbreth} (Accessed: January 2023), [Online]

\bibitem{imagenet}
{ImageNet}. Available:\url{https://image-net.org/download.php/} (Accessed: January 2023), [Online]

\bibitem{Pneumonia}
{Pneumonia | Disease or Condition of the Week | CDC}. Available: \url{https://bit.ly/35aOE7Q} (Accessed: January 2023), [Online]

\bibitem{who}
{WHO Coronavirus (COVID-19) Dashboard}. Available: \url{https://covid19.who.int/} (Accessed: January 2023), [Online]

\bibitem{wearablecons}
{Aditi Pai}: {Survey: One third of wearable device owners stopped using them within six months}. Available: \url{https://bit.ly/3yjuzrC} (Accessed: January 2023), [Online]

\bibitem{al2009load}
Al~Amin, M.T., Barua, S., Vhaduri, S., Rahman, A.: {Load aware broadcast in mobile ad hoc networks}. In: IEEE International Conference on Communications (ICC) (2009)

\bibitem{amoh2015deepcough}
Amoh, J., Odame, K.: Deepcough: A deep convolutional neural network in a wearable cough detection system. In: 2015 IEEE Biomedical Circuits and Systems Conference (BioCAS). pp.~1--4. IEEE (2015)

\bibitem{cai2023discovering}
Cai, J., Vhaduri, S., Luo, X.: Discovering covid-19 coughing and breathing patterns from unlabeled data using contrastive learning with varying pre-training domains. INTERSPEECH  (2023)

\bibitem{chang2022covnet}
Chang, Y., Jing, X., Ren, Z., Schuller, B.W.: Covnet: A transfer learning framework for automatic covid-19 detection from crowd-sourced cough sounds. Frontiers in Digital Health  \textbf{3},  799067 (2022)

\bibitem{chen2020estimating}
Chen, C.Y., Vhaduri, S., Poellabauer, C.: {Estimating sleep duration from temporal factors, daily activities, and smartphone use}. In: IEEE Computer Society Computers, Software, and Applications Conference (COMPSAC) (2020)

\bibitem{cheung2020context}
Cheung, W., Vhaduri, S.: {Context-Dependent Implicit Authentication for Wearable Device Users}. In: IEEE International Symposium on Personal, Indoor, and Mobile Radio Communications (PIMRC) (2020)

\bibitem{cheung2020continuous}
Cheung, W., Vhaduri, S.: {Continuous Authentication of Wearable Device Users from Heart Rate, Gait, and Breathing Data}. In: IEEE RAS \& EMBS International Conference on Biomedical Robotics and Biomechatronics (BioRob) (2020)

\bibitem{cohen2020novel}
Cohen-McFarlane, M., Goubran, R., Knoefel, F.: Novel coronavirus cough database: Nococoda. Ieee Access  \textbf{8},  154087--154094 (2020)

\bibitem{dibbo2023sok}
Dibbo, S.V.: Sok: Model inversion attack landscape: Taxonomy, challenges, and future roadmap. In: 2023 IEEE 36th Computer Security Foundations Symposium (CSF). pp. 439--456. IEEE (2023)

\bibitem{dibbo2021onphone}
Dibbo, S.V., Cheung, W., Vhaduri, S.: {On-Phone CNN Model-based Implicit Authentication to Secure IoT Wearables}. In: EAI International Conference on Safety and Security in Internet of Things (SaSeIoT) (2021)

\bibitem{dibbo2021effect}
Dibbo, S.V., Kim, Y., Vhaduri, S.: {Effect of Noise on Generic Cough Models}. In: IEEE International Conference on Wearable and Implantable Body Sensor Networks (BSN) (2021)

\bibitem{dibbo2021visualizing}
Dibbo, S.V., Kim, Y., Vhaduri, S., Poellabauer, C.: Visualizing college students’ geo-temporal context-varying significant phone call patterns. In: 2021 IEEE 9th International Conference on Healthcare Informatics (ICHI). pp. 381--385. IEEE (2021)

\bibitem{dibbo2023lcanets++}
Dibbo, S.V., Moore, J.S., Kenyon, G.T., Teti, M.A.: Lcanets++: Robust audio classification using multi-layer neural networks with lateral competition. arXiv preprint arXiv:2308.12882  (2023)

\bibitem{drugman2013objective}
Drugman, T., Urbain, J., Bauwens, N., Chessini, R., Valderrama, C., Lebecque, P., Dutoit, T.: Objective study of sensor relevance for automatic cough detection. IEEE journal of biomedical and health informatics  \textbf{17}(3),  699--707 (2013)

\bibitem{elfaramawy2018wireless}
Elfaramawy, T., Fall, C.L., Arab, S., Morissette, M., Lellouche, F., Gosselin, B.: A wireless respiratory monitoring system using a wearable patch sensor network. IEEE Sensors Journal  \textbf{19}(2),  650--657 (2018)

\bibitem{gemmeke2017audio}
Gemmeke, J.F., Ellis, D.P., Freedman, D., Jansen, A., Lawrence, W., Moore, R.C., Plakal, M., Ritter, M.: Audio set: An ontology and human-labeled dataset for audio events. In: 2017 IEEE international conference on acoustics, speech and signal processing (ICASSP). pp. 776--780. IEEE (2017)

\bibitem{hui2021wearable}
Hui, X., Zhou, J., Sharma, P., Conroy, T.B., Zhang, Z., Kan, E.C.: Wearable rf near-field cough monitoring by frequency-time deep learning. IEEE Transactions on Biomedical Circuits and Systems  \textbf{15}(4),  756--764 (2021)

\bibitem{imran2020ai4covid}
Imran, A., Posokhova, I., Qureshi, H.N., Masood, U., Riaz, M.S., Ali, K., John, C.N., Hussain, M.I., Nabeel, M.: Ai4covid-19: Ai enabled preliminary diagnosis for covid-19 from cough samples via an app. Informatics in Medicine Unlocked  \textbf{20},  100378 (2020)

\bibitem{kamei2022use}
Kamei, T., Kanamori, T., Yamamoto, Y., Edirippulige, S.: The use of wearable devices in chronic disease management to enhance adherence and improve telehealth outcomes: a systematic review and meta-analysis. Journal of Telemedicine and Telecare  \textbf{28}(5),  342--359 (2022)

\bibitem{kim2020understanding}
Kim, Y., Vhaduri, S., Poellabauer, C.: {Understanding College Students' Phone Call Behaviors Towards a Sustainable Mobile Health and Wellbeing Solution}. In: International Conference on Systems Engineering (2020)

\bibitem{laguarta2020covid}
Laguarta, J., Hueto, F., Subirana, B.: Covid-19 artificial intelligence diagnosis using only cough recordings. IEEE Open Journal of Engineering in Medicine and Biology  \textbf{1},  275--281 (2020)

\bibitem{liaqat2021coughwatch}
Liaqat, D., Liaqat, S., Chen, J.L., Sedaghat, T., Gabel, M., Rudzicz, F., de~Lara, E.: Coughwatch: Real-world cough detection using smartwatches. In: ICASSP 2021-2021 IEEE International Conference on Acoustics, Speech and Signal Processing (ICASSP). pp. 8333--8337. IEEE (2021)

\bibitem{lien2023challenges}
Lien, C.W., Vhaduri, S.: Challenges and opportunities of biometric user authentication in the age of iot: A survey. ACM Computing Surveys  \textbf{55}(12),  1--39 (2023)

\bibitem{mcfee2015librosa}
McFee, B., Raffel, C., Liang, D., Ellis, D.P., McVicar, M., Battenberg, E., Nieto, O.: librosa: Audio and music signal analysis in python. In: Proceedings of the 14th python in science conference. vol.~8, pp. 18--25 (2015)

\bibitem{meng2019speech}
Meng, H., Yan, T., Yuan, F., Wei, H.: Speech emotion recognition from 3d log-mel spectrograms with deep learning network. IEEE access  \textbf{7},  125868--125881 (2019)

\bibitem{muratyan2021opportunistic}
Muratyan, A., Cheung, W., Dibbo, S.V., Vhaduri, S.: {Opportunistic Multi-Modal User Authentication for Health-Tracking IoT Wearables}. In: EAI International Conference on Safety and Security in Internet of Things (SaSeIoT) (2021)

\bibitem{orlandic2021coughvid}
Orlandic, L., Teijeiro, T., Atienza, D.: The coughvid crowdsourcing dataset, a corpus for the study of large-scale cough analysis algorithms. Scientific Data  \textbf{8}(1),  1--10 (2021)

\bibitem{pahar2021covid}
Pahar, M., Klopper, M., Warren, R., Niesler, T.: Covid-19 cough classification using machine learning and global smartphone recordings. Computers in Biology and Medicine  \textbf{135},  104572 (2021)

\bibitem{pahar2021machine}
Pahar, M., Niesler, T.: Machine learning based covid-19 detection from smartphone recordings: cough, breath and speech. arXiv pre-print  (2021)

\bibitem{piczak2015esc}
Piczak, K.J.: Esc: Dataset for environmental sound classification. In: Proceedings of the 23rd ACM international conference on Multimedia. pp. 1015--1018 (2015)

\bibitem{richards2022economic}
Richards, F., Kodjamanova, P., Chen, X., Li, N., Atanasov, P., Bennetts, L., Patterson, B.J., Yektashenas, B., Mesa-Frias, M., Tronczynski, K., et~al.: Economic burden of covid-19: A systematic review. ClinicoEconomics and Outcomes Research: CEOR  \textbf{14}, ~293 (2022)

\bibitem{sharma2020coswara}
Sharma, N., Krishnan, P., Kumar, R., Ramoji, S., Chetupalli, S.R., Ghosh, P.K., Ganapathy, S., et~al.: Coswara--a database of breathing, cough, and voice sounds for covid-19 diagnosis. arXiv preprint arXiv:2005.10548  (2020)

\bibitem{sharmin2015visualization}
Sharmin, M., Raij, A., Epstien, D., Nahum-Shani, I., Beck, J.G., Vhaduri, S., Preston, K., Kumar, S.: Visualization of time-series sensor data to inform the design of just-in-time adaptive stress interventions. In: Proceedings of the 2015 ACM International Joint Conference on Pervasive and Ubiquitous Computing. pp. 505--516 (2015)

\bibitem{simhadri2022understanding}
Simhadri, S., Vhaduri, S.: {Understanding User Trust in Different Recommenders and Smartphone Applications}. In: EAI International Conference on Wireless Mobile Communication and Healthcare (MobiHealth) (2022)

\bibitem{simonyan2014very}
Simonyan, K., Zisserman, A.: Very deep convolutional networks for large-scale image recognition. arXiv preprint arXiv:1409.1556  (2014)

\bibitem{stojanovic2020headset}
Stojanovi{\'c}, R., {\v{S}}kraba, A., Lutovac, B.: A headset like wearable device to track covid-19 symptoms. In: 2020 9th Mediterranean Conference on Embedded Computing (MECO). pp.~1--4. IEEE (2020)

\bibitem{vhaduri2020nocturnal}
Vhaduri, S.: {Nocturnal cough and snore detection using smartphones in presence of multiple background-noises}. In: ACM SIGCAS Conference on Computing and Sustainable Societies (COMPASS) (2020)

\bibitem{vhaduri2014estimating}
Vhaduri, S., Ali, A., Sharmin, M., Hovsepian, K., Kumar, S.: {Estimating drivers' stress from GPS traces}. In: International Conference on Automotive User Interfaces and Interactive Vehicular Applications (AutomotiveUI) (2014)

\bibitem{vhaduri2019towards}
Vhaduri, S., Brunschwiler, T.: {Towards automatic cough and snore detection}. In: IEEE International Conference on Healthcare Informatics (ICHI) (2019)

\bibitem{vhaduri2023bag}
Vhaduri, S., Cheung, W., Dibbo, S.V.: {Bag of On-Phone ANNs to Secure IoT Objects Using Wearable and Smartphone Biometrics}. IEEE Transactions on Dependable and Secure Computing  \textbf{20}(3),  1--12 (2023)

\bibitem{vhaduri2023predicting}
Vhaduri, S., Cho, J., Meng, K.: {Predicting Unreliable Response Patterns in Smartphone Health Surveys: A Case Study with the Mood Survey}. Elsevier Smart Health Journal  \textbf{28},  100398 (2023)

\bibitem{vhaduri2022predicting}
Vhaduri, S., Dibbo, S.V., Chen, C.Y.: Predicting a user's demographic identity from leaked samples of health-tracking wearables and understanding associated risks. In: 2022 IEEE 10th International Conference on Healthcare Informatics (ICHI). IEEE (2022)

\bibitem{vhaduri2021predicting}
Vhaduri, S., Dibbo, S.V., Chen, C.Y., Poellabauer, C.: {Predicting Next Call Duration: A Future Direction to Promote Mental Health in the Age of Lockdown}. In: IEEE Computer Society Computers, Software, and Applications Conference (COMPSAC) (2021)

\bibitem{vhaduri2021HIAuth}
Vhaduri, S., Dibbo, S.V., Cheung, W.: {HIAuth: A Hierarchical Implicit Authentication System for IoT Wearables Using Multiple Biometrics}. IEEE Access  \textbf{9},  116395--116406 (2021)

\bibitem{vhaduri2023implicit}
Vhaduri, S., Dibbo, S.V., Cheung, W.: Implicit iot authentication using on-phone ann models and breathing data. Elsevier Internet of Things  \textbf{24} (2023)

\bibitem{vhaduri2021deriving}
Vhaduri, S., Dibbo, S.V., Kim, Y.: {Deriving College Students’ Phone Call Patterns to Improve Student Life}. IEEE Access  \textbf{9},  96453--96465 (2021)

\bibitem{vhaduri2023environment}
Vhaduri, S., Dibbo, S.V., Kim, Y.: {Environment Knowledge-Driven Generic Models to Detect Coughs from Audio Recordings}. IEEE Open Journal of Engineering in Medicine and Biology  \textbf{4},  1--12 (2023)

\bibitem{vhaduri2016assessing}
Vhaduri, S., Munch, A., Poellabauer, C.: {Assessing health trends of college students using smartphones}. In: IEEE Healthcare Innovation Point-of-Care Technologies Conference (HI-POCT) (2016)

\bibitem{vhaduri2015design}
Vhaduri, S., Poellabauer, C.: {Design and Implementation of a Remotely Configurable and Manageable Well-being Study}. In: EAI SWIT-Health (2015)

\bibitem{vhaduri2016cooperative}
Vhaduri, S., Poellabauer, C.: {Cooperative discovery of personal places from location traces}. In: International Conference on Computer Communication and Networks (ICCCN) (2016)

\bibitem{vhaduri2016human}
Vhaduri, S., Poellabauer, C.: {Human factors in the design of longitudinal smartphone-based wellness surveys}. In: IEEE International Conference on Healthcare Informatics (ICHI) (2016)

\bibitem{vhaduri2017design}
Vhaduri, S., Poellabauer, C.: {Design factors of longitudinal smartphone-based health surveys}. Journal of Healthcare Informatics Research  \textbf{1}(1),  52--91 (2017)

\bibitem{vhaduri2017towards}
Vhaduri, S., Poellabauer, C.: Towards reliable wearable-user identification. In: 2017 IEEE International Conference on Healthcare Informatics (ICHI) (2017)

\bibitem{vhaduri2017wearable}
Vhaduri, S., Poellabauer, C.: {Wearable device user authentication using physiological and behavioral metrics}. In: IEEE International Symposium on Personal, Indoor, and Mobile Radio Communications (PIMRC) (2017)

\bibitem{vhaduri2018biometric}
Vhaduri, S., Poellabauer, C.: {Biometric-based wearable user authentication during sedentary and non-sedentary periods}. International Workshop on Security and Privacy for the Internet-of-Things (IoTSec)  (2018)

\bibitem{vhaduri2018hierarchical}
Vhaduri, S., Poellabauer, C.: {Hierarchical cooperative discovery of personal places from location traces}. IEEE Transactions on Mobile Computing  \textbf{17}(8),  1865--1878 (2018)

\bibitem{vhaduri2018impact}
Vhaduri, S., Poellabauer, C.: {Impact of different pre-sleep phone use patterns on sleep quality}. In: IEEE International Conference on Wearable and Implantable Body Sensor Networks (BSN) (2018)

\bibitem{vhaduri2018opportunisticICHI}
Vhaduri, S., Poellabauer, C.: {Opportunistic discovery of personal places using smartphone and fitness tracker data}. In: IEEE International Conference on Healthcare Informatics (ICHI) (2018)

\bibitem{vhaduri2019multi}
Vhaduri, S., Poellabauer, C.: {Multi-Modal Biometric-Based Implicit Authentication of Wearable Device Users}. IEEE Transactions on Information Forensics and Security  \textbf{14}(12),  3116--3125 (2019)

\bibitem{vhaduri2019summary}
Vhaduri, S., Poellabauer, C.: {Summary: Multi-modal Biometric-based Implicit Authentication of Wearable Device Users}. arXiv preprint arXiv:1907.06563  (2019)

\bibitem{vhaduri2018opportunisticTBD}
Vhaduri, S., Poellabauer, C.: {Opportunistic discovery of personal places using multi-source sensor data}. IEEE Transactions on Big Data  \textbf{7}(2),  383--396 (2021)

\bibitem{vhaduri2017discovering}
Vhaduri, S., Poellabauer, C., Striegel, A., Lizardo, O., Hachen, D.: {Discovering places of interest using sensor data from smartphones and wearables}. In: IEEE Ubiquitous Intelligence \& Computing (UIC) (2017)

\bibitem{vhaduri2020adherence}
Vhaduri, S., Prioleau, T.: {Adherence to personal health devices: A case study in diabetes management}. In: EAI International Conference on Pervasive Computing Technologies for Healthcare (PervasiveHealth) (2020)

\bibitem{vhaduri2022understanding}
Vhaduri, S., Simhadri, S.: {Understanding User Concerns and Choice of App Architectures in Designing Audio-based mHealth Apps}. Elsevier Smart Health Journal  \textbf{26},  100341 (2022)

\bibitem{vhaduri2019nocturnal}
Vhaduri, S., Van~Kessel, T., Ko, B., Wood, D., Wang, S., Brunschwiler, T.: {Nocturnal cough and snore detection in noisy environments using smartphone-microphones}. In: IEEE International Conference on Healthcare Informatics (ICHI) (2019)

\end{thebibliography}

\end{document}